\documentstyle[12pt]{article}
\begin{document}

\vspace{0.5cm}

\begin{center}
{\bf{GEOMETRIC PROPERTIES OF QCD STRING FROM WILLMORE
FUNCTIONAL}}

\vspace{0.5cm}

R.Parthasarathy{\footnote{e-mail
address:sarathy@imsc.ernet.in}} \\ 
The Institute of Mathematical Sciences \\
C.I.T.Campus, Taramani Post \\
Chennai 600 113, India. \\
and \\
K.S.Viswanathan{\footnote{e-mail address:kviswana@sfu.ca}} \\
Department of Physics \\
Simon Fraser University \\
Burnaby \\
Canada, V5A 1S6. \\  
\end{center}

\vspace{1.0cm}

{\noindent{\it{Abstract}}}

The extremum of the Willmore-like functional for
$m$-dimensional Riemannian surface immersed in $d$-dimensional
Riemannian manifold under normal variations is studied 
and various cases of
interest are examined. This study is used to relate the
parameters of QCD string action, including the
Polyakov-Kleinert extrinsic curvature action, with the
geometric properties of the world sheet. The world sheet has
been shown to have {\it{negative stiffness}} on the basis of
the geometric considerations. 

\vspace{0.5cm}

{\noindent{MSC indices:
51P05;53C42;53C80;81V05}}

\vspace{0.5cm}


\newpage 

\noindent{\bf{I.Introduction}}

\vspace{0.5cm}

QCD strings is a string theory in 4-dimensions. It has been
realized by Polyakov [1] and independently by Kleinert [2]
that for QCD strings, added extrinsic curvature action to the
usual Nambu-Goto (NG) area term is appropriate. In particular
the theory with extrinsic curvature action alone has been
shown to be asymptotically free [1,2,3] - a feature relevant
to describe QCD. By considering the 1-loop multi-instanton
effects in the theory of 2-dimensional world sheet in $R^3$
and $R^4$, the grand partition function has been found to be
that of a 2-dimensional modified Coulomb gas system with long
range order in the infra-red region and, in plasma phase in the
ultra-violet region [4]. The above result uses the running
coupling constant and the string world sheet is stable against
small fluctuations along the normal (transverse) directions in
the infra-red region and avoids crumpling. Thus the
Polyakov-Kleinert string provides a relevant
description of colour flux tubes between quarks in QCD.

\vspace{0.5cm}

In order to remove the unphysical ghost poles and to realize a
lowest energy state, Kleinert and Chervyakov [5] recently
proposed a {\it new} string model with {\it negative} sign for
the extrinsic curvature action i.e., they {\it hypothesize}
negative {\it stiffness} for the gluonic flux tubes, inspired
by properties of magnetic flux tubes in Type-II superconductor
and of Nielsen-Olesen vortices in relativistic gauge models.
They propose an effective string action as,  
\begin{eqnarray}
S&=&\frac{(c-1)}{2}M^2\int d^2\xi \sqrt{g} g^{\alpha\beta}
{\nabla}_{\alpha}X^{\mu}(\xi)\frac{1}{c-e^{
{\nabla}^2/{\mu}^2}}
{\nabla}_{\beta}X^{\mu}(\xi), \nonumber 
\end{eqnarray}
where $X^{\mu}(\xi)$ ; $\mu\ =\ 1,2,3,4$ are the worldsheet
coordinates, ${\xi}_1,{\xi}_2$ are the local isothermal
coordinates on the surface, $g_{\alpha\beta}\ =\
{\partial}_{\alpha}X^{\mu}{\partial}_{\beta}X^{\mu}$ is the
induced metric (first fundamental form) on the world sheet,
${\nabla}_{\alpha}$ is the covariant derivative on the
surface, $M$ is dimensionfull (mass dimension) constant and
$c$ is a dimensionless constant. The propagator from the
quadratic part in $X^{\mu}$, in momentum space [5], is 
\begin{eqnarray}
G(k^2)&=&\frac{1}{(c-1)}\ \frac{c-e^{-k^2/{\mu}^2}}{k^2},
\nonumber 
\end{eqnarray}
and for small momentum, this is
\begin{eqnarray}
G(k^2)&\simeq & \frac{1+k^2/{\Lambda}^2}{k^2}, \nonumber 
\end{eqnarray}
with ${\Lambda}^2\ =\ (c-1)/{\mu}^2$. This has a single pole
at $k^2\ =\ 0$ with negative stiffness ${\alpha}_0\ =\
-{\Lambda}^2/M^2$, in contrast to the propagator in
Polyakov-Kleinert model $1/(k^2(1+k^2/{\Lambda}^2)$, which has
unphysical pole at $k^2\ =\ -{\Lambda}^2$ and which has
positive stiffness of ${\Lambda}^2/M^2$. Approximating the
full propagator by its low momentum expression, Kleinert and
Chervyakov [5] have    
proposed an action at low momentum region as,
\begin{eqnarray}
S_{KC} &=& \frac{1}{2}M^2 \int d^2\xi\ \sqrt{g}
\ g^{\alpha\beta} {\nabla}_{\alpha}X^{\mu}\ \frac{1}{1\ -\
\frac{{\nabla}^2}{{\Lambda}^2}}\ {\nabla}_{\beta}X^{\mu}. 
\end{eqnarray}
Such an
action (1) has the
the high temperature behaviour as that of large-N QCD [6]. 
The {\it negative} extrinsic curvature term can be seen from
(1) by expanding the non-local term, using the Gauss equation
\begin{eqnarray}
{\nabla}_{\alpha}{\nabla}_{\beta}X^{\mu} &=& 
H^i_{\alpha\beta}N^{i\mu},
\end{eqnarray}
where $i\ =\ 1,2$ and $H^{i}_{\alpha\beta}$ are the components
of the extrinsic curvature (second fundamental form) along the
two normals $N^{i\mu}$ to the world sheet, and the Weingarten
equation [4]
\begin{eqnarray}
{\nabla}_{\alpha}N^{i\mu} &=&
-H^{i\gamma}_{\alpha}{\partial}_{\gamma}X^{\mu},
\end{eqnarray}
where the covariant derivative ${\nabla}_{\alpha}$ in (3)
incorporates the connection in the normal frame as well. 
By expanding $(1\ -\ \frac{{\nabla}^2}
{{\Lambda}^2})^{-1}$ in (1) and
realizing $X^{\mu}$ is a scalar on the world sheet, (1) can be
written as
\begin{eqnarray}
S_{KC} &=& \frac{1}{2} M^2 \int d^2\xi\ \sqrt{g}\ \{2\ +\
\frac{1}{{\Lambda}^2}g^{\alpha\beta}{\partial}_{\alpha}X^{\mu}
\  g^{\gamma\delta}{\nabla}_{\gamma}
({\nabla}_{\delta}{\partial}_{\beta}
X^{\mu}) - \cdots \}, \nonumber
\end{eqnarray}
where we have retained up to the $\frac{1}{{\Lambda}^2}$ 
term for
illustration. Upon using (2) and (3) and the fact
${\partial}_{\alpha}X^{\mu}\ N^{i\mu}\ =\ 0$, the above
expression simplifies to
\begin{eqnarray}
S_{KC} &=& \frac{1}{2} M^2 \int d^2\xi\ \sqrt{g}\ \{2 \ -\
\frac{1}{{\Lambda}^2} H^{i\alpha\beta}H^i_{\alpha\beta}\}.
\nonumber 
\end{eqnarray}
But then,
\begin{eqnarray}
H^{i\alpha\beta}H^i_{\alpha\beta} &=& 4{\mid H\mid}^2 + R,
\nonumber
\end{eqnarray}
where ${\mid H\mid}^2\ =\ H^iH^i$, with $H^i\ =\
\frac{1}{2}g^{\alpha\beta}H^i_{\alpha\beta}$ and $R$ is the
scalar curvature of the world sheet. In view of this, the
expression for $S_{KC}$ becomes
\begin{eqnarray}
S_{KC} &\simeq & M^2 \int \sqrt{g}\ d^2\xi -
\frac{2M^2}{{\Lambda}^2}\int \sqrt{g}\ {\mid H\mid}^2\ d^2\xi
- \frac{M^2}{{\Lambda}^2}\int \sqrt{g}\ R\ d^2\xi + \cdots 
\end{eqnarray}
In above $M^2$ plays the role of string tension. The extrinsic
curvature action (the second term in (4)) has {\it negative}
stiffness. The third term is just the Euler characteristic of
the surface which is a topological invariant action. It is
clear from (1) that unphysical poles can be avoided by
appealing to surfaces with negative stiffness.

\vspace{0.5cm}

It will be worthwhile to examine whether the negative
stiffness is favoured from purely geometric considerations of
the surface. In this context, the Willmore surfaces which
extremize the Willmore functional [7]  
\begin{eqnarray}
S_W &=& \int \sqrt{g}\ {\mid H\mid}^2\ d^2\xi, \nonumber 
\end{eqnarray}
become relevant. It is the purpose of this paper to first  
consider general Willmore functional for
$m$-dimensional surface immersed in $d$-dimensional Riemannian
space ($m\ <\ d$) and study various cases of interest.  
Then using the results, {\it{we compare the classical equation of
motion for (4) with immersion in flat space, with that of the
Willmore functional for immersion in a Riemannian space}},
thereby showing the effects of the Nambu-Goto term in (4)
could be accounted for by considerations of the Willmore
functional in a curved space. 

\vspace{0.5cm}

The extremum of Willmore functional for hypersurfaces in
Euclidean space ($E^3$) has been dealt with in detail by
Willmore [7] and, by Chen [8] for $m$-dimensional oriented
closed hypersurface in Euclidean space $E^{m+1}$. Willmore and
Jhaveri [9] extended to $m$-dimensional manifold immersed as a
hypersurface of a general ($m+1$)-dimensional Riemannian
manifold and Weiner [10] to that of 2-dimensional surface in a
general Riemannian manifold. 
In this paper, we examine the Willmore functional for the
general case of $m$-dimensional Riemann surface immersed in
$d$-dimensional Riemann space and then consider various cases
of interest.  
As an application of this study,  
we will compare the equation of motion of Willmore functional
for $2$-dimensional surface immersed in $d$-dimensional space
with the classical equation of motion of QCD string  
to relate the QCD string parameters,
namely, the string tension and stiffness parameter to the
geometrical properties of the surface.   

\vspace{1.0cm}

\noindent{\bf{II. EFFECTS DUE TO NORMAL VARIATIONS}}

\vspace{0.5cm}

For an $m$-dimensional surface $\Sigma $ immersed in a
$d$-dimensional ($d\ >\ m$) Riemannian manifold {$\Sigma'$} with
metric $h_{\mu\nu}\ ;\ (\mu,\nu\ =\ 1,2,\cdots d)$, we have the
induced metric on $\Sigma$ as
\begin{eqnarray}
g_{\alpha\beta} &=& {\partial}_{\alpha}X^{\mu}\
{\partial}_{\beta}X^{\nu}\ h_{\mu\nu},
\end{eqnarray}
where the indices $\alpha,\beta$ take values $1,2,\cdots m$
and $X^{\mu}\ =\ X^{\mu}({\xi}_1,{\xi}_2\cdots ,{\xi}_m)$,
with ${\xi}_{\alpha}$'s as coordinates on $\Sigma$. There are
$(d-m)$ unit normals at a point $P\in \Sigma$, denoted by
$N^{i\mu};\ (i\ =\ 1,2,\cdots (d-m))$, chosen to satisfy  
\begin{eqnarray}
N^{i\mu}\ N^{j\nu}\ h_{\mu\nu} &=& {\delta}^{ij} \nonumber \\
{\partial}_{\alpha}X^{\mu}\ N^{i\nu}\ h_{\mu\nu} &=& 0, \ \ \
\forall \ i\ =\ 1,2,\cdots (d-m);\ \ \forall \ \alpha\ =\ 1,2,
\cdots ,m.
\end{eqnarray}
Repeated indices will be appropriately summed over in this
paper. The equation of Gauss [11] for $\Sigma$,
\begin{eqnarray}
{\nabla}_{\alpha}{\nabla}_{\beta}X^{\mu} &\equiv &
{\partial}_{\alpha}{\partial}_{\beta}X^{\mu}-{\Gamma}^{\gamma}
_{\alpha\beta}{\partial}_{\gamma}X^{\mu}+{\tilde{\Gamma}}^
{\mu}_{\nu\rho}{\partial}_{\alpha}X^{\nu}{\partial}_{\beta}
X^{\rho} \nonumber \\
& = & H^i_{\alpha\beta}N^{i\mu},
\end{eqnarray}
defines the {\it{second fundamental form}} $H^i_{\alpha\beta}$.
${\Gamma}^{\gamma}_{\alpha\beta}$ and
${\tilde{\Gamma}}^{\mu}_{\nu\rho}$ are the connections on
$\Sigma$ and $\Sigma'$ determined by $g_{\alpha\beta}$ and
$h_{\mu\nu}$ respectively. The $(d-m)$ normals $N^{i\mu}$
satisfy the Weingarten equation
\begin{eqnarray}
{\nabla}_{\alpha}N^{i\mu} &\equiv&
{\partial}_{\alpha}N^{i\mu}+{\tilde{\Gamma}}^{\mu}_{\nu\rho}
{\partial}_{\alpha}X^{\nu}N^{i\rho}-A^{ij}_{\alpha}N^{j\mu}
\nonumber \\
&=& -H^{i\beta}_{\alpha}{\partial}_{\beta}X^{\mu},
\end{eqnarray}
where $A^{ij}_{\alpha}\ =\
N^{j\nu}({\partial}_{\alpha}N^{i\mu}+{\tilde{\Gamma}}^{\mu}_
{\nu\rho}{\partial}_{\alpha}X^{\nu}N^{i\rho})h_{\mu\nu}$ is
the $m$-dimensional gauge field or connection in the normal
bundle. We need the Gauss equation [11]
\begin{eqnarray}
{\tilde{R}}_{\mu\nu\rho\sigma}{\partial}_{\alpha}X^{\mu}
{\partial}_{\beta}X^{\nu}{\partial}_{\gamma}X^{\rho}
{\partial}_{\delta}X^{\sigma} &=& R_{\alpha\beta\gamma
\delta}+H^{i}_{\beta\gamma}H^i_{\alpha\delta}-H^i_
{\beta\delta}H^i_{\alpha\gamma},
\end{eqnarray}
where $R_{\alpha\beta\gamma\delta}$ and ${\tilde{R}}_{\mu\nu
\rho\sigma}$ are the Riemann symbols of the first kind for
$\Sigma$ and $\Sigma'$ respectively. We introduce the mean
curvature $H^i$ (there are ($d-m$) such quantities) by
\begin{eqnarray}
H^i &=& \frac{1}{m} g^{\alpha\beta}H^i_{\alpha\beta}.
\end{eqnarray}     

\vspace{0.5cm}

The variations of the surface can be described by the
variations of $X^{\mu}({\xi}_1,\cdots {\xi}_m)$ as
$X^{\mu}({\xi}_1,\cdots {\xi}_m)\ +\ \delta X^{\mu}
({\xi}_1,\cdots {\xi}_m)$. In general $\delta
X^{\mu}\ =\ {\phi}^i N^{i\mu}\ +\ {\partial}_{\alpha}X^{\mu}
{\eta}^{\alpha}$, comprising of variations along 
$m$ tangent directions and $(d-m)$ normals. The tangential 
variations
 are related to the
structure equations [12]. So, we consider only the normal
variations and accordingly,
\begin{eqnarray}
\delta X^{\mu} &=& {\phi}^i N^{i\mu}.
\end{eqnarray} 
Using (5) it follows for normal variations
\begin{eqnarray}
\delta \sqrt{g} &=& -m\sqrt{g} {\phi}^i H^i, \nonumber \\
\delta g^{\alpha\beta} &=& 2{\phi}^i H^{i\alpha\beta}.
\end{eqnarray}
From (10) it follows
\begin{eqnarray}
\delta H^i &=& \frac{2}{m} {\phi}^k H^{k\alpha\beta}
H^i_{\alpha\beta} + \frac{1}{m} g^{\alpha\beta}\delta H^i_{
\alpha\beta},
\end{eqnarray}
using the second equation in (12). For hypersurfaces, there
will be only one normal and in such a case, the computation of
$\delta H$ has been given in Ref.9. 
The evaluation of $\delta H^i_{\alpha\beta}$
for 2-dimensional surface in $d$-dimensional Riemannian
manifold is described in Ref.12. The computation of $\delta
H^i_{\alpha\beta}$ for $m$-dimensional surface in
$d$-dimensional Riemannian manifold is involved and we give
here the relevant steps for the sake of completeness.

\vspace{0.5cm}

From (7) we have
\begin{eqnarray}
H^i_{\alpha\beta} &=& \{{\partial}_{\alpha}{\partial}_{\beta}
X^{\mu} + {\tilde{\Gamma}}^{\mu}_{\rho\sigma}{\partial}
_{\alpha}X^{\rho}{\partial}_{\beta}X^{\sigma}\}N^{i\nu}h_{
\mu\nu}.
\end{eqnarray}
Using (6), we have
\begin{eqnarray}
\delta N^{j\mu}\ N^{i\nu}\ h_{\mu\nu} + N^{j\mu}\ \delta
N^{i\nu}\ h_{\mu\nu} &=& -N^{j\mu}N^{i\nu}\ \delta h_{\mu\nu},
\nonumber \\
 & & \nonumber \\
{\partial}_{\gamma}(\delta X^{\mu}) N^{i\nu}\ h_{\mu\nu} +
{\partial}_{\gamma}X^{\mu}\ \delta N^{i\nu}\ h_{\mu\nu}
&=& - {\partial}_{\gamma}X^{\mu} N^{i\nu}\ \delta h_{\mu
\nu}, \nonumber
\end{eqnarray}
and then,
\begin{eqnarray}
g^{\alpha\beta}H^i\ \delta H^i_{\alpha\beta} &=&
g^{\alpha\beta}H^i{\tilde{R}}_{\rho\sigma\nu\lambda}{\partial}
_{\alpha}X^{\nu}{\partial}_{\beta}X^{\rho}\ \delta X^{
\sigma} N^{i\lambda} \nonumber  \\
 &+& g^{\alpha\beta}H^i({\nabla}_{\alpha}{\nabla}_{\beta}\ 
\delta X^{\mu})\ N^i_{\mu} 
+\frac{m}{2}H^iH^j N^{j\mu} N^{i\nu}\ ({\partial}_
{\lambda}h_{\mu\nu})\ \delta X^{\lambda} \nonumber \\  
&-& m H^iH^j{\tilde{
\Gamma}}^{\mu}_{\rho\lambda}N^{j\sigma}\ \delta X^{\lambda}
\ N^i_{\mu}.
\end{eqnarray}
The last two terms cancel each other after expanding
${\tilde{\Gamma}}^{\mu}_{\rho\lambda}$ and using $i\
\leftrightarrow j$ symmetry. 
Now using (11) and (8),
we find
\begin{eqnarray}
g^{\alpha\beta}H^i\ \delta H^i_{\alpha\beta} &=&
g^{\alpha\beta}H^i{\tilde{R}}_{\rho\sigma\nu\lambda}{\partial}
_{\alpha}X^{\nu}{\partial}_{\beta}X^{\rho}{\phi}^k N^{k
\sigma}N^{i\lambda} \nonumber \\
&+& H^k({\nabla}_{\alpha}{\nabla}^{\alpha}{\phi}^k) - H^i
H^i_{\alpha\beta}H^{k\alpha\beta}{\phi}^k.
\end{eqnarray} 

\vspace{0.5cm}

We consider the extremum of the following functional
\begin{eqnarray}
W &=& \int \sqrt{g}\ \left( H^iH^i\right)^{\frac{m}{2}}\
d^m\xi,
\end{eqnarray}
which reduces to Willmore functional for $m\ =\ 2$ and to that
of Chen [8] for $m$-dimensional {\it hypersurface} as well
with Willmore and Jhaveri [9] for $m$-dimensional {\it
hypersurface} in $(m+1)$ dimensional Riemannian manifold. The
normal variations of (17) give the equations of motion. 
Taking the normal variations of (17) and
using (12), (13) and (16), we obtain
\begin{eqnarray}
\delta W &=& \int \sqrt{g} (H^jH^j)^{\frac{m}{2}-1}H^k
({\nabla}_{\alpha}{\nabla}^{\alpha}{\phi}^k)\ d^m\xi \nonumber
\\
&-& m\int \sqrt{g}{\phi}^kH^k(H^jH^j)^{\frac{m}{2}}\ d^m\xi
\nonumber \\
&+& \int \sqrt{g}(H^jH^j)^{\frac{m}{2}-1}{\phi}^k H^iH^i_
{\alpha\beta}H^{k\alpha\beta}\ d^m\xi \nonumber \\
&+&\int \sqrt{g} (H^jH^j)^{\frac{m}{2}-1}H^i g^{\alpha\beta}
{\tilde{R}}_{\rho\sigma\nu\lambda}{\partial}_{\alpha}X^{\nu}
{\partial}_{\beta}X^{\rho}{\phi}^k N^{k\sigma}N^{i\lambda}\
 d^m\xi.
\end{eqnarray} 
Equating this to zero and using
\begin{eqnarray}
\int \ \sqrt{g}\ (H^jH^j)^{\frac{m}{2}-1}\ H^k\
({\nabla}_{\alpha}{\nabla}^{\alpha}{\phi}^k)\ d^m\xi &=&
\nonumber \\ 
\hspace{2.0cm} \int \ \sqrt{g}\ {\phi}^k\   
 {\nabla}_{\alpha}{\nabla}^
{\alpha}\left( (H^jH^j)^{\frac{m}{2}-1} H^k\right)\ d^m\xi,
\end{eqnarray}
we obtain the equation of motion for (17) as 
\begin{eqnarray}
{\nabla}_{\alpha}{\nabla}^{\alpha}\left(
(H^jH^j)^{\frac{m}{2}-1}H^k\right)-mH^k(H^jH^j)^{\frac{m}{2}}
+(H^jH^j)^{\frac{m}{2}-1}H^iH^i_{\alpha\beta}H^{k\alpha\beta}
\nonumber \\
+(H^jH^j)^{\frac{m}{2}-1}H^i g^{\alpha\beta}{\tilde{R}}_
{\rho\sigma\nu\lambda}{\partial}_{\alpha}X^{\nu}{\partial}_
{\beta}X^{\rho}\ N^{k\sigma}\ N^{i\lambda}\ =\ 0,
\end{eqnarray}
since (18) must hold for all allowed ${\phi}^k$.  

\vspace{0.5cm}

We now consider various cases.

\vspace{1.0cm}

{\noindent{\it{Case.1$\ $ hypersurface in Euclidean space}}}

\vspace{0.5cm}

Let $\Sigma $ be an $m$-dimensional hepersurface in $d\ =\
m+1$ dimensional {\it{Euclidean}} space, i.e., ${\Sigma}'\ =\
E^{m+1}$. As there will be only one normal for a hypersurface,
$H^jH^j\ =\ H^2$ and (20) reduces to
\begin{eqnarray}
{\nabla}_{\alpha}{\nabla}^{\alpha}(H^{m-1}) - mH^{m+1} +
H^{m-1} H_{\alpha\beta}H^{\alpha\beta} &=& 0,
\end{eqnarray} 
and in this case the Gauss equation (9) when contracted with
$g^{\alpha\gamma}g^{\beta\delta}$ gives
\begin{eqnarray}
H^{\alpha\beta}H_{\alpha\beta} &=&-R + m^2H^2, \nonumber 
\end{eqnarray}
where $R$ is the curvature scalar of the hypersurface $\Sigma
$. Then (21) becomes
\begin{eqnarray}
{\nabla}_{\alpha}{\nabla}^{\alpha}(H^{m-1})  
+ m(m-1)H^{m+1} - H^{m-1}R
&=& 0,
\end{eqnarray}
which is the result of Chen [8] and agrees with Eqn.5.59 of
Willmore [7]. 

\vspace{0.5cm}

{\noindent{\it{Case.2$\ $ hypersurface in Riemannian space}}}

\vspace{0.5cm}

Let $\Sigma $ be an $m$-dimensional {\it{hypersurface}}
immersed in $d\ =\ m+1$ dimensional {\it{Riemannian}} manifold
$\Sigma '$. Then Eqn.20 becomes
\begin{eqnarray}
{\nabla}_{\alpha}{\nabla}^{\alpha}H^{m-1} - mH^{m+1} +
H^{m-1}H_{\alpha\beta}H^{\alpha\beta} \nonumber \\
 + H^{m-1}g^{\alpha\beta}{\tilde{R}}_{\rho\sigma
\nu\lambda}{\partial}_{\alpha}X^{\nu}{\partial}_{\beta}X^
{\rho}N^{\sigma}N^{\lambda}\ =\ 0. 
\end{eqnarray}
In this case, the equation of Gauss (9) is
\begin{eqnarray}
R_{\alpha\beta\gamma\delta}&=&H_{\beta\delta}H_{\alpha\gamma}-
H_{\beta\gamma}H_{\alpha\delta}+{\tilde{R}}_{\mu\nu\rho
\sigma}{\partial}_{\alpha}X^{\mu}{\partial}_{\beta}X^{\nu}
{\partial}_{\gamma}X^{\rho}{\partial}_{\delta}X^{\sigma},
\nonumber 
\end{eqnarray}
which when contracted with $g^{\alpha\gamma}g^{\beta\delta}$
gives
\begin{eqnarray}
R&=&-H^{\alpha\beta}H_{\alpha\beta}+m^2H^2+{\tilde{R}}_{\mu
\nu\rho\sigma}{\partial}_{\alpha}X^{\mu}{\partial}_{\beta}
X^{\nu}{\partial}_{\gamma}X^{\rho}{\partial}_{\delta}X^
{\sigma}g^{\alpha\gamma}g^{\beta\delta}.
\end{eqnarray} 
The completeness relation 
found in Ref.12 will now be used and it is
\begin{eqnarray}
h^{\mu\nu} &=&
g^{\alpha\beta}{\partial}_{\alpha}X^{\mu}{\partial}_{\beta}X^
{\nu} + \sum^{d-2}_{i}N^{i\mu}N^{i\nu}. 
\end{eqnarray}
For hypersurfaces, (25) is simply
\begin{eqnarray}
h^{\mu\nu} &=&
g^{\alpha\beta}{\partial}_{\alpha}X^{\mu}{\partial}_{\beta}
X^{\nu} + N^{\mu}N^{\nu},
\end{eqnarray}
and using this in (24), we find
\begin{eqnarray}
H^{\alpha\beta}H_{\alpha\beta} &=& -R + m^2H^2 + \tilde{R}- 
2{\tilde{R}}_{\mu\nu}N^{\mu}N^{\nu}.
\end{eqnarray}
Using (26) in the last
term of (23), we have
\begin{eqnarray}
g^{\alpha\beta}{\tilde{R}}_{\rho\sigma\nu\lambda}{\partial}_
{\alpha}X^{\nu}{\partial}_{\beta}X^{\rho}N^{\sigma}N^
{\lambda}&=& {\tilde{R}}_{\rho\sigma\nu\lambda}h^{\rho\nu}
N^{\sigma}N^{\lambda}-{\tilde{R}}_{\rho\sigma\nu\lambda}
N^{\nu}N^{\sigma}N^{\rho}N^{\lambda} \nonumber \\
&=& {\tilde{R}}_{\sigma\lambda}N^{\sigma}N^{\lambda},
\nonumber
\end{eqnarray}
and so the Eqn.23 becomes,
\begin{eqnarray}
{\nabla}_{\alpha}{\nabla}^{\alpha}H^{m-1}+m(m-1)H^{m+1}+
H^{m-1}\{-R+\tilde{R}-{\tilde{R}}_{\mu\nu}N^{\mu}N^{\nu}
\}&=& 0.
\end{eqnarray}  

\vspace{0.5cm}

We analyse (28), by
choosing an orthogonal frame at $P\ \in \ \Sigma$ such that
the matrix $H^{\alpha\beta}$ is diagonal. Then
$H^{\alpha\beta}H_{\alpha\beta}\ =\ \sum^{m}_{i=1}h^{2}_i$,
and $m^2H^2\ =\ (\sum^m_{i=1}h_i)^2$. 
(See [9]). Then, Eqn.28 can
be written as,
\begin{eqnarray}
{\nabla}_{\alpha}{\nabla}^{\alpha}H^{m-1} &=&
-H^{m-1}\{\sum^{m}_{i=1} h^2_i - \frac{1}{m}(\sum^m_{i=1}h)^2
+{\tilde{R}}_{\mu\nu}N^{\mu}N^{\nu}\},
\end{eqnarray}
since (27) gives
\begin{eqnarray}
-R+{\tilde{R}}+m^2H^2-2{\tilde{R}}_{\mu\nu}N^{\mu}N^{\nu}&=&
\sum^m_{i=1}h^2_i. \nonumber
\end{eqnarray}
It is to be noted that $\sum^m_{i=1}h^2_i
-\frac{1}{m}(\sum^m_{i=1}h_i)^2\ \geq \ 0$. For ${\tilde{R}}
_{\mu\nu}$ positive-definite, it is seen from (29) that
${\nabla}_{\alpha}{\nabla}^{\alpha}H^{m-1}$ has the same sign
as $-H^{m-1}$.  

\vspace{0.5cm}

{\noindent{\it{Case.3$\ $ 2-dimensional surface in Riemannian
space}}}

\vspace{0.5cm}

Let $\Sigma$ be a 2-dimensional surface immersed in
${\Sigma}'$. Then (20) becomes,
\begin{eqnarray}
{\nabla}_{\alpha}{\nabla}^{\alpha}H^k-2H^k(H^jH^j)+H^iH^i
_{\alpha\beta}H^{k\alpha\beta}+ \nonumber \\ 
H^ig^{\alpha\beta}{\tilde{
R}}_{\rho\sigma\nu\lambda}{\partial}_{\alpha}X^{\nu}
{\partial}_{\beta}X^{\rho}N^{k\sigma}N^{i\lambda}&=&0,
\end{eqnarray}
agreeing with Eqn.35 of Ref.12.

\vspace{0.5cm}

{\noindent{\it{Case.4$\ $ Immersions in space of constant
curvature}}}

\vspace{0.5cm}

Consider $\Sigma'$ space to be a space of constant curvature
i.e., de-Sitter or anti-de-Sitter type. In this case [13]
\begin{eqnarray}
{\tilde{R}}_{\mu\nu\rho\sigma}&=&
K(h_{\mu\rho}h_{\nu\sigma}-h_{\mu\sigma}h_{\nu\rho}).
\end{eqnarray}
Then,
\begin{eqnarray}
{\tilde{R}}_{\nu\sigma}\ =\
h^{\mu\rho}{\tilde{R}}_{\mu\nu\rho\sigma}&=&K(d-1)h_{\nu
\sigma}, \nonumber \\
 & & \nonumber \\
\tilde{R} &=& h^{\mu\sigma}{\tilde{R}}_{\mu\sigma}\ =\
Kd(d-1).
\end{eqnarray}
Then (20) becomes,
\begin{eqnarray}
{\nabla}_{\alpha}{\nabla}^{\alpha}\left(
(H^jH^j)^{\frac{m}{2}-1}H^k\right)+mH^k(H^jH^j)^{\frac{m}{2}-1
}(K-H^{\ell}H^{\ell}) \nonumber \\
+(H^jH^j)^{\frac{m}{2}-1}H^iH^i_{\alpha\beta}H^{k\alpha\beta}\
\ =\ 0. 
\end{eqnarray}

\vspace{0.3cm}

Similarly Eqn.28 in Case.2, becomes,
\begin{eqnarray}
{\nabla}_{\alpha}{\nabla}^{\alpha}H^{m-1}+m(m-1)H^{m+1}
-R H^{m-1}&=&0,
\end{eqnarray}
for $m$-dimensional hypersurface immersed in $\Sigma'$ space
of constant curvature. 

\vspace{0.3cm}

Eqn.30 of Case.3 i.e., 2-dimensional surface immersed in
$\Sigma'$, a space of constant curvature, becomes,
\begin{eqnarray}
{\nabla}_{\alpha}{\nabla}^{\alpha}H^k - 2H^k(H^jH^j) + H^i
H^{i\alpha\beta}H^k_{\alpha\beta} + 2KH^k &=& 0.
\end{eqnarray}

In spite of the simplifications, these equations are still  
difficult to solve
explicitly without further choices for the geometry. 

\vspace{1.0cm}

{\noindent{\bf{III. QCD STRING AND WILLMORE FUNCTIONAL}}}

\vspace{0.5cm}

As explained in the Introduction, it appears that a candidate
action for describing QCD string has to involve the extrinsic
geometry of the world sheet, regarded as a 2-dimensional
Riemannian surface immersed in $R^4$. With {\it{negative}}
stiffness, Kleinert and Chervyakov [5] successfully obtained
the correct hight temperature behaviour as in large-N QCD [6].
Further evidence for the role of the extrinsic geometry in QCD
stems from the $U(N);\ \ N\ \rightarrow \ \infty$ lattice
gauge theory calculations of Kazakov [14], Kostov [15] and
O'Brien and Zuber [16]. These calculations confirm the
equivalence of multicolour QCD and string theory in which the
resulting surfaces intersect at self-intersections. It is
known that the self-intersection number involves extrinsic
geometry. We consider the action (4) without the Euler
characteristic term, as
\begin{eqnarray}
S_{KC} &=& T \int \sqrt{g}\ d^2\xi + {\alpha}_0 \int
\sqrt{g}\ {\mid H\mid}^2\ d^2\xi,
\end{eqnarray}
where $T$ is the string tension, ${\alpha}_0$ a measure of
stiffness of the QCD string immersed in $R^d$ (say) and $H^2\
=\ H^iH^i;\ i=1,2 \cdots (d-2)$. The extremum of (36) can be
easily found using (12), (13), and (16) for normal variations.
We find the equation of motion for (36) as
\begin{eqnarray}
{\nabla}_{\alpha}{\nabla}^{\alpha}H^k-\frac{2T}{{\alpha}_0}H^k
-2H^kH^jH^j+H^iH^{k\alpha\beta}H^i_{\alpha\beta}&=&0,
\end{eqnarray}
for the stiffness parameter ${\alpha}_0\ \neq\ 0$. This
non-linear equation is complicated and it will be wothwhile
hence to draw some information from this. 

Kholodenko and
Nesterenko [17] proposed an approach in this direction by
considering (36) for immersion in $R^3$ and relating to the
extremum of the Willmore functional for immersion in $S^3$. We
will generalize this approach here, by relating (37) to (20)
for $m\ =\ 2$, which is (30). Eqn.30 has the same form as (37)
provided we identify,
\begin{eqnarray}
-\frac{2T}{{\alpha}_0}H^k &=&
H^i{\tilde{R}}_{\rho\sigma\nu\lambda}{\partial}_{\alpha}X^{\nu
}{\partial}_{\beta}X^{\rho}N^{k\sigma}N^{i\lambda}g^{\alpha
\beta}.
\end{eqnarray}
Upon using (25), this becomes,
\begin{eqnarray}
-\frac{2T}{{\alpha}_0}H^k&=&
H^i({\tilde{R}}_{\sigma\lambda}N^{k\sigma}N^{i\lambda}- 
{\tilde{R}}_{\rho\sigma\nu\lambda}N^{j\nu}N^{j\rho}N^{k\sigma}
N^{i\lambda}),
\end{eqnarray}
a {\it{new relation}} among the string tension, stiffness
parameter, mean curvature scalar and the geometric properties
of ${\Sigma}'$. In order to make (39) manageable, we take
${\Sigma}'$ a space of constant curvature as in (31) and (32).
Then it can be seen,
\begin{eqnarray}
{\tilde{R}}_{\sigma\lambda}N^{k\sigma}N^{i\lambda}&=&K(d-1)
{\delta}_{ik}, \nonumber \\
 & & \nonumber \\
{\tilde{R}}_{\rho\sigma\nu\lambda}N^{j\nu}N^{j\rho}N^{k\sigma}
N^{i\lambda}&=&K(d-3){\delta}^{ik},
\end{eqnarray}
and so (39) becomes
\begin{eqnarray}
-\frac{2T}{{\alpha}_0}H^k &=& K(d-1)H^k-K(d-3)H^k\ =\ 2KH^k.
\end{eqnarray}
Now as we have assumed that $H^k$'s are not zero, it follows
\begin{eqnarray}
\frac{T}{{\alpha}_0} &=& -K.
\end{eqnarray}
It is noted here that the dimensionality of ${\Sigma}'$ does
not directly appear in relating $T$, the string tension, and
${\alpha}_0$, the stiffness parameter, with $K$. From this, it
follows that the stiffness parameter can be positive for $K<0$
(Anti-de-Sitter background) or negative for $K>0$ (de-Sitter
background).    

\vspace{1.0cm}

{\noindent{\bf{VI. CONCLUSIONS}}}

\vspace{0.5cm}

The extremum of the Willmore functional for $m$-dimensional
surface immersed in $d$-dimensional Riemannian space is
studied under the normal variations of the immersed surface.
Various cases of interest are examined. In particular the
equation of motion for a 2-dimensional surface immersed in
spaces of constant curvature, is compared with the equation of
motion of the Polyakov-Kleinert action of the QCD string
considered as a Riemann surface immersed in $R^4$, to obtain a
new relation connecting the string tension and stiffness
parameter of the QCD string on the one hand and the constant
$K$ of the Riemann space (31). This relation $T/{\alpha}_0\
=\ -K$, is independent of the dimensionality of ${\Sigma}'$.
For positive $K$, favoured by positive-definteness of
${\tilde{R}}_{\mu\nu}$ (see Case.2) from (32), it follows that
negative stiffness is recommended by geometric considerations.
This result agrees with the observation of Kleinert and
Chervyakov [5] using (physical) QCD string.   
Thus the QCD
string world sheet regarded as a 2-dimensional surface
immersed in $R^4$ has been shown to favour negative stiffness by
comparing its classical equation with that of a Willmore
2-dimensional surface immersed in a space of constant
curvature.   

\vspace{1.0cm}

{\noindent{\bf {REFERNCES}}}

\vspace{0.5cm}

\begin{enumerate}

\item A.M.Polyakov, Nucl.Phys. {\bf B268} (1986) 406 ; {\bf
B486} (1997) 23. 

\item H.Kleinert, Phys.Lett. {\bf B174} (1986) 335; \\
  Phys.Rev.Lett. {\bf 58} (1987) 1915; Phys.Lett. {\bf B189}
(1987) 187.

\item K.S.Viswanathan, R.Parthasarathy and D.Kay, \\
      Ann.Phys. (N.Y), {\bf 206} (1991) 237.

\item K.S.Viswanathan and R.Parthasarathy, Phys.Rev. {\bf D51}
(1995) 5830.

\item H.Kleinert and A.M.Chervyakov, Phys.Lett. {\bf B381}
(1996) 286. \\
      See also, H.Kleinert, Phys.Lett. {\bf B211} (1988) 151.

\item J.Polchinski, Phys.Rev.Lett. {\bf 68} (1992) 1267.

\item T.J.Willmore, {\it Total Curvature in Riemannian
Geometry}, \\
      Ellis Harwood, 1982.

\item B.Y.Chen, J.London.Math.Soc. (2) {\bf 6} (1973) 321.

\item T.J.Willmore and C.S.Jhaveri, Quart.J.Math.Oxford. (2)
{\bf 23} (1972) 319.

\item J.L.Weiner, Indiana Univ. Math. Journal. {\bf 27} (1978)
19.

\item L.P.Eisenhart, {\it {Riemannian Geometry}}, \\
      Princeton University Press. 1966.

\item K.S.Viswanathan and R.Parthasarathy, Phys.Rev. {\bf D55}
(1997) 3800. 

\item S.Weinberg, {\it{Gravitation and Cosmology}}, \\
      John-Wiley and Sons.Inc.N.Y. 1972.

\item V.A.Kazakov, Phys.Lett. {\bf B128} (1983) 316. 

\item I.K.Kostov, Phys.Lett. {\bf B138} (1984) 191. 

\item K.H.O'Brien and J.B.Zuber, Nucl.Phys. {\bf B253} (1985)
621. 

\item A.L.Kholodenko and V.V.Nesterenko, J.Geom.Phys. {\bf 16}
(1995) 15. 
\end{enumerate}
\end{document}